# Integrating Deep Learning into CAD/CAE System: Generative Design and Evaluation of 3D Conceptual Wheel

Soyoung Yoo[1]

Sunghee Lee[1]

Seongsin Kim[1]

Kwang Hyeon Hwang[2]

Jong Ho Park[2]

Namwoo Kang[3,*]

[1]Department of Mechanical Systems Engineering,
Sookmyung Women's University, 04310, Seoul, Korea

[2]Hyundai Motor Company, 445706, Hwaseong-Si, Gyeonggi-Do, Korea

[3]The Cho Chun Shik Graduate School of Green Transportation,
Korea Advanced Institute of Science and Technology, 34141, Daejeon, Korea

*Corresponding author: nwkang@kaist.ac.kr

**Abstract**

Engineering design research integrating artificial intelligence (AI) into computer-aided design (CAD) and computer-aided engineering (CAE) is actively being conducted. This study proposes a deep learning-based CAD/CAE framework in the conceptual design phase that automatically generates 3D CAD designs and evaluates their engineering performance. The proposed framework comprises seven stages: (1) 2D generative design, (2) dimensionality reduction, (3) design of experiment in latent space, (4) CAD automation, (5) CAE automation, (6) transfer learning, and (7) visualization and analysis. The proposed framework is demonstrated through a road wheel design case study and indicates that AI can be practically incorporated into an end-use product design project. Engineers and industrial designers can jointly review a large number of generated 3D CAD models by using this framework along with the engineering performance results estimated by AI and find conceptual design candidates for the subsequent detailed design stage.

# 1. Introduction

Deep learning, which is a part of artificial intelligence (AI) technology and learns meaningful patterns based on deep neural network structures from large amounts of data, demonstrates remarkable performances in various areas (LeCun et al., 2015). In recent years, the expectations for deep learning research have increased in computer-aided design (CAD) and computer-aided engineering (CAE), which are the core of new product development (Guo et al., 2016; Umetani, 2017; Zhang et al., 2018; Cunningham et al., 2019; Khadilkar et al., 2019; Oh et al., 2019; Williams et al., 2019; Nie et al., 2020). CAD/CAE methods can be easily combined with deep learning in comparison with other engineering fields. This condition is because many deep learning studies have been conducted using CAD data for classification and segmentation (Maturana and Scherer, 2015; Qi et al., 2017), and CAE research has long used machine learning to build metamodels (Wang & Shan, 2007).

The conceptual design phase requires a surrogate model (or metamodel) that can quickly evaluate the engineering performance of a large number of design candidates. Deep learning is considered a powerful method for surrogate modeling in the field of CAE research due to its ability to approximate high-dimensional and highly nonlinear physics (Cunningham et al., 2019; Khadilkar et al., 2019; Du et al., 2020a), and hyperparameter search affects the accuracy and robustness of surrogate models (Du et al., 2020b). However, the largest problem is that engineers have to first create a large number of CAD models and collect CAE results to utilize deep learning.

Generative design can be used as a solution to overcome the limitation of data shortage. Generative design refers to computational design methods that can automatically conduct design exploration under constraints defined by designers (Shea et al., 2005; Krish 2011; Singh & Gu, 2012; Jang et al., 2021). Recent generative design research utilizes topology optimization and deep learning to enable exploration of large design spaces effectively and efficiently (Oh et al., 2018; Oh et al., 2019; Jang et al., 2021; Kallioras & Lagaros, 2020; Sun & Ma, 2020).

In a previous study, Oh et al. (2019) proposed a deep learning-based generative design method for 2D wheel design. This study aims to extend this generative design method to 3D wheel design problem for an industrial application to demonstrate its feasibility in the automotive industry. This study proposes a deep learning-based CAD/CAE framework by combining generative design, CAD/CAE automation, and deep learning technologies. The proposed framework is specifically design for the conceptual design phase, and its purpose is to automatically generate 3D CAD data and evaluate them through deep learning to find feasible conceptual designs in the early design phase.

The proposed framework includes (1) 2D generative design, (2) dimensionality reduction, (3) design of experiment (DOE) in latent space, (4) 3D CAD automation, (5) CAE automation, (6) transfer learning, and (7) visualization and analysis. Using the proposed framework in the conceptual wheel design process, engineering designers can automatically generate a large number of feasible 3D wheel CAD models, and immediately predict modal analysis result based on a disk-view 2D wheel design without a 3D CAD/CAE modeling process.

The proposed framework is expected to change the conceptual design phase of the vehicle wheel. The existing design process is inefficient because the industrial designer creates only a small portion of the design, whereas the engineer creates a CAD/CAE model and provides feedback to the industrial designer, thereby repeating the design modification process. However, AI first generates a large amount of CAD models using the proposed framework and evaluates them in terms of engineering performance metrics. Industrial designers and engineers can then select good conceptual design candidates and modify them for detailed designs. Industrial designers can instantly evaluate the engineering performance of new designs without the need for engineers to review them. Owing to AI, designers and engineers can collaborate efficiently.

The remainder of this paper is organized as follows. Section 2 introduces the related research. Section 3 introduces the proposed deep CAD/CAE framework. Sections 4 and 5 provide the details of the proposed framework. Section 6 presents the discussion, conclusion, and future research directions.

# 2. Related Work

## 2.1. CAD/CAE research with deep learning

In computer science, deep learning research using 3D CAD data being actively conducted. The key to learning 3D CAD data is a pipeline design that preprocesses high-dimensional CAD data to be used as input into deep learning architectures. The three most widely used types of deep learning models based on the preprocessing technology are as follows: (1) voxel-based models (e.g., VoxNet), (2) point cloud-based models (e.g., PointNet), and (3) view-based model (e.g., multiview convolutional neural networks (CNNs)). VoxNet (Maturana & Schiller, 2015) uses a 3D CNN architecture that takes voxels representing 3D CAD as input. In a discontinuous 3D space, voxel data have a value of 1 when the object occupies the grid, otherwise they have a value of 0. PointNet (Qi et al., 2017) learns the features of points directly from 3D CAD. Point clouds are defined as a number of points ($x$, $y$, $z$) on the surface of a 3D object, representing the data as a set of coordinates. Multiview based models (e.g., Su et al., 2015, Kanezaki et al., 2018) use 2D images rather than 3D information, where the 2D images of a 3D CAD are captured with a virtual camera around it. In addition to the three main types, a method of learning the mesh data of the CAD model exists, that is, MeshNet (Feng et al., 2019).

On the basis of the 3D deep learning models using 3D CAD data introduced above, engineering design research has focused on CAE research predicting engineering performance by learning the data obtained through the finite element method (FEM). Cunningham et al. (2019) conducted a study to predict the aerodynamic performance by preprocessing a 3D aircraft model into 3D point clouds and applying deep learning. Khadilkar et al. (2019) automatically generated 3D CAD models of various shapes and then sliced the models to obtain cross sections. A CNN model has been proposed to predict stresses on a cross section that occur during bottom-up 3D printing using both the cross section and 3D point cloud of the 3D model. In vehicle system design, researchers have used an autoencoder to parameterize automotive 3D mesh models, generate 3D mesh models of various shapes, and learn the computational fluid dynamics (CFD) results of models with Gaussian process regression analysis (Umetani, 2017; Umetani & Bickel, 2018). Computer aided manufacturing (CAM) research is also in progress. Williams et al. (2019) proposed a 3D CNN to estimate manufacturability by predicting the part mass, support material mass, and build time required to three-dimensionally print a CAD model. Zhang et al. (2018) conducted a study to automatically generate CAD data according to machining features, learn data through a 3D CNN, and predict machining features. Compared with 3D design studies, more studies have been performed to investigate the engineering performances of 2D designs. Guo et al. (2016) proposed a CNN model that approximates a steady flow on a vehicle's side view 2D geometry, whereas Nie et al. (2020) proposed a generative model that predicts the 2D stress field of a 2D cantilever bar by learning the design space and boundary conditions.

The bottleneck of deep learning-based 3D CAD/CAE research is to collect a large amount of 3D CAD data in the target product area. Therefore, we need to first discuss how to generate data efficiently for deep learning research.

## 2.2. Generative design

Generative design refers to computational design methods that can automatically conduct design exploration under constraints defined by designers (Shea et al., 2005; Krish 2011; Singh & Gu, 2012; Jang et al., 2021). Generative design can provide initial designs and new inspiration in the conceptual design phase (Kallioras & Lagaros, 2020). Conventional generative design controls feasible shape changes through geometric parameterization of CAD models and uses a variety of exploration methods, such as genetic algorithms (GAs), to generate different designs (Krish 2011; Singh & Gu, 2012).

Topology optimization (Bendsoe & Kikuchi, 1988), which optimizes material layout (density) within a given design space to find the optimal (stiffest) design, can be used for design exploration. Jang et al. (2021) summarize the idea of exploring diverse designs through topology optimization as follows: The first is to find various local optima for the same problem. Diverse designs can be found through various initial designs, optimizers, and filtering methods (Andreassen et al., 2011). The second is to find a Pareto set for a multiobjective (disciplinary) optimization problem. For practical use in the industry, topology optimization must meet different requirements simultaneously (Kunakote & Bureerat, 2011). The third one is to diversify the definition of topology optimization problems. New designs can be generated by varying design conditions, such as load and boundary conditions, material selection, and manufacturing methods (Matejka et al., 2018).

The industry is rebranding topology optimization as "generative design," providing CAD tools that leverage topology optimization for design exploration, and efforts to apply it to real-world product development are accelerating (Autodesk, 2020a). However, these CAD tools does not use deep learning.

In academia, deep learning research on improving the design exploration performance of topology optimization-based generative design is in the early stage (Oh et al., 2018; Oh et al., 2019; Jang et al., 2021; Kallioras & Lagaros, 2020; Sun & Ma, 2020). Kallioras and Lagaros (2020) integrated reduced order models and convolution filters of deep belief networks to generate various topology designs. Sun and Ma (2020) proposed reinforcement learning-based generative design that does not need preoptimized topology data. Deep generative design (Oh et al., 2019), which is the basis of this study, generates diverse designs similar to the actual designs in the market through the integration of topology optimization and generative model of deep learning. Traditional topology optimization only considers the engineering point of view, resulting in a design that is usually unfamiliar to customers and designers. However, deep generative design solves the multiobjective problem of minimizing compliance and maximizing similarity to market reference designs, allowing a designer to generate designs that look like real products on sale. Jang et al. (2021) enhanced the diversity of deep generative design by applying reinforcement learning.

### 2.3. Deep learning used in the proposed framework

Three main deep learning techniques, namely, CNN, autoencoder, and transfer learning, are used in the proposed framework. As supervised learning, deep neural networks (DNNs) and CNNs are mainly used to build surrogate models of engineering problems, as introduced in Section 2.1. In particular, CNNs are frequently used in fields where computer vision is applied and show high performance in learning shape patterns to recognize images and objects (Krizhevsky et al., 2012). CNN architecture consists of a combination of convolutional and pooling layers, adding fully connected layers on top of them. We used CNN for building a surrogate model for modal analysis in the proposed framework.

As unsupervised learning, autoencoders are primarily used for dimensional reduction (Hinton and Salakhutdinov, 2006). Autoencoders can compress high-dimensional input data into a low-dimensional latent space. In the autoencoder architecture, the size of the input layer is the same as the size of the output layer. The network that compresses the input data into the latent space is called an encoder, and the network that restores the latent space to the output data is called the decoder. We used a convolutional autoencoder for dimensionality reduction of CAD data; this autoencoder consists of only convolutional and pooling layers (Masci et al., 2011).

Transfer learning methods are widely used to overcome the lack of training data in the target domain (Pan and Yang, 2009). Transfer learning facilitates learning by applying pretrained feature extractor models with large datasets (in the source domain) to problems with small related datasets (in the target domain) by supplementing insufficient information. The pretrained autoencoder's encoder can be used as a feature extractor, and we can perform transfer learning by adding a fully connected layer on top of the encoder and fine-tuning the model (Cunningham et al., 2019). We used this approach to improve the predictive performance of our surrogate model.

## 3. Deep CAD/CAE Framework

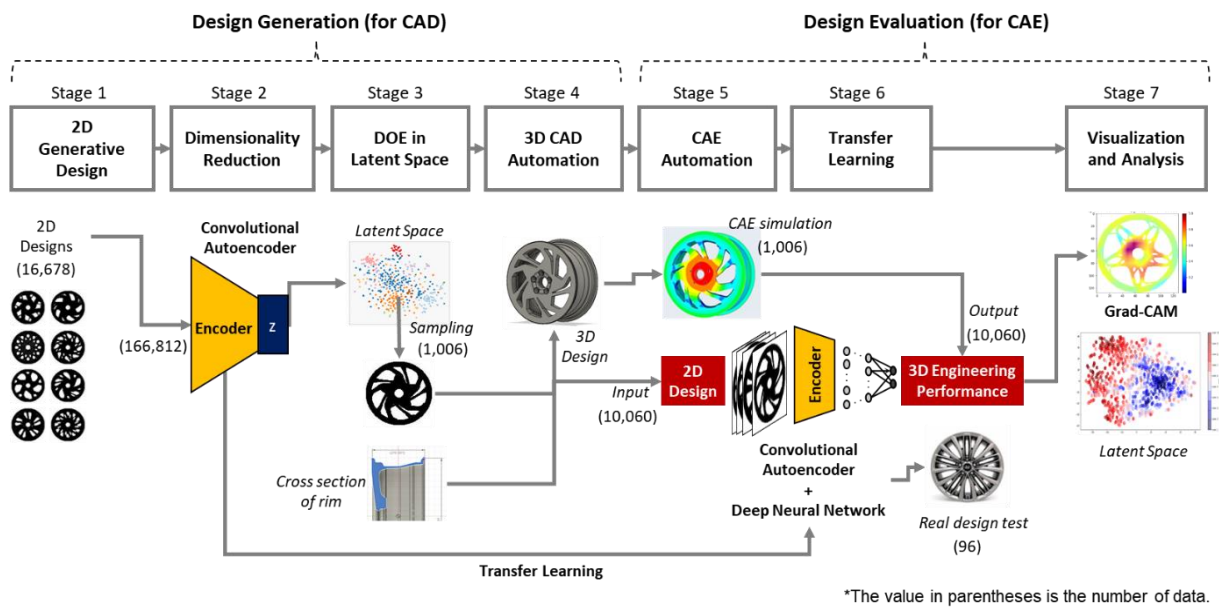

**Figure 1. Deep CAD/CAE framework**

The deep learning-based CAD/CAE framework proposed herein comprises seven stages, and the framework is presented in Figure 1. The goals of the proposed framework are summarized as follows: The first goal is to develop a fully automated CAD process that generates 3D CAD data based on 2D generative design (stages 1 to 4). The second goal is to develop a deep learning model to evaluate the engineering performance of a 3D CAD model using 2D design as input and 3D CAE simulation result as output (stages 5 to 7). In the early stages of product development, we can generate and evaluate the numerous conceptual designs using the proposed framework.

The description of each stage of the proposed framework is presented as follows:

- *Stage 1. 2D generative design*: To generate various 2D wheel designs, we adopted our previous deep generative design process (Oh et al. 2019), which is an effective method combining topology optimization and deep learning to create many engineered structure designs. First, we collected the image data of a commercial wheel as reference designs; subsequently, based on these reference designs, a number of new topology designs of 2D disk-view wheels were generated automatically. In this study, we created 16,689 2D wheel designs, and more details pertaining to Stage 1 are provided in Section 3.

- *Stage 2. Dimensionality reduction*: In this stage, the dimensions of the 2D wheel design generated in Stage 1 are reduced. Dimensionality reduction helps to overcome the curse of dimensionality, thereby allows us to extract important features from 2D wheel designs. In this study, we used a convolutional autoencoder (Masci et al., 2011), which performs well in reducing the dimensionality of images. In our research, 128 × 128 2D images were mapped into a 128-dimensional latent space. Subsequently, the trained encoder of the convolutional autoencoder was used in Stages 3 and 6. Details of Stage 2 are provided in Section 4.

- *Stage 3. DOE in latent space*: This stage pertains to the DOE process for drawing 2D wheel design samples from the latent space, which are then used for creating CAD data. Because this latent space comprises the feature vectors of the wheel design, the data distribution is more meaningful than the original high-dimensional space. In this study, Latin hypercube sampling (LHS) (Viana, 2016) was used, and 1,030 2D wheel designs were sampled from the latent space. A detailed description of Stage 3 is provided in Section 5.

- *Stage 4. 3D CAD automation*: In this stage, 3D CAD data to be used as input to CAE are automatically generated. First, the 2D wheel design undergoes preprocessing, which involves four steps: (1) smoothing and sharpening of edges, (2) edge extraction, (3) conversion of edges into coordinate data, and (4) grouping of edge coordinates. Next, the process of generating a 3D CAD based on the 2D wheel image and the cross-section image of the given rim is automated. In our study, Autodesk Fusion 360 (Autodesk, 2020b) was used to automate 3D CAD modeling. The details of Stage 4 are provided in Section 6.

- *Stage 5. CAE automation*: In this stage, CAE simulation data are collected using the 3D CAD data generated in Stage 4. In this study, we conducted modal analysis to verify the natural frequency of the lateral mode, and the result was stored as labeled data used for deep learning. Altair SimLab (Altair, 2020) was used for CAE automation. The details of Stage 5 are provided in Section 7.

- *Stage 6. Transfer learning*: In this stage, a surrogate model is built to predict the CAE simulation results by using a CNN with transfer learning. Using the 2D wheel design as input, the deep learning model predicts the natural frequency and mass as the output. In this study, to solve the problem of insufficient data, data augmentation and transfer learning were conducted by combining a DNN with the encoder of a convolutional autoencoder, which was pretrained in Stage 2. Furthermore, we used an ensemble technique to reduce overfitting and improve the prediction performance. Stage 6 is described in detail in Section 8.

- *Stage7. Visualization and analysis*: In this stage, CAD/CAE engineers can visualize and explain the deep learning results to gain new insights into and evaluate the reliability of the results. The latent space created in Stage 2 can be visualized in two dimensions to examine the relationship between the wheel shape and natural frequency. In addition, the wheel shape that significantly affects the natural frequency can be identified by Grad-CAM. More details regarding Stage 7 are provided in Section 9.

# 4. Design Generation (Stages 1 to 4)

## 4.1. 2D Generative Design (Stage 1)

Stage 1 involves creating large amounts of 2D disk-view wheel designs. A part of the deep generative design framework was used in this study, which combined deep learning and topology optimization, as suggested by Oh et al. (2019).

Topology optimization is typically used for structural design, in which a design area is divided into elements and the optimal material density of the elements is determined to minimize compliance considering a given load and boundary condition. Oh et al. (2019) added a term to the typical topology optimization problem, which minimizes the distance from a reference design to a new topology design. Consequently, the problem becomes a multiobjective problem to obtain a new topology design that is similar to the reference design and has a low compliance simultaneously. The proposed optimization formulation is as follows:

$$\begin{aligned}
\min \quad & f(\mathbf{x}) = \mathbf{U}^\mathrm{T}\mathbf{K}(\mathbf{x})\mathbf{U} + \lambda\|\mathbf{x}_r - \mathbf{x}\|_1 \\
\text{s.t.} \quad & \frac{V(\mathbf{x})}{V_0} = f \\
& \mathbf{KU} = \mathbf{F} \\
& 0 \leq x_e \leq 1, \ e = 1, \ldots, N_e
\end{aligned} \quad (1)$$

where $x_e$ is the design variable representing the density of element $e$, and $\mathbf{x}$ is the density vector. $\mathbf{U}$ is the displacement vector, $\mathbf{K}$ is the global stiffness matrix, and $\mathbf{U}^\mathrm{T}\mathbf{K}(\mathbf{x})\mathbf{U}$ corresponds to the compliance. $\mathbf{x}_r$ indicates the reference design, $\|\cdot\|_1$ indicates the L1 norm, and $\lambda$ is the similarity weight. The larger the $\lambda$, the closer the optimal design is to the reference design. For a small $\lambda$, the reference design is disregarded, and it is optimized to minimize the compliance. $f$ is the volume fraction, $V(\mathbf{x})$ is the volume of the material, and $V_0$ is the volume of the design domain.

In topology optimization, length scale control and geometric feature control can be used in applying manufacturing constraints. These methods can be grouped into four classes, namely, filtering techniques, projection mapping, parameter space reduction, and direct imposition of constraints (Sutradhar et al., 2017). Our study uses density filtering (Bruns and Tortorelli, 2001; Bourdin, 2001) with Heaviside projection (Guest et al., 2004; Sigmund, 2007). Detailed equations and optimization methods are available in Oh et al. (2019).

In topology optimization, a single optimal design is obtained when the designer defines the objective function, loads, and boundary conditions. A generative design involves creating multiple optimal designs by varying the weights of multiobjective and design parameters, as explained in Section 2.2. The generative design process is shown in Figure 2. In this study, three design parameters were defined. (1) similarity weights ($\lambda$) with five levels (i.e., 0.0005, 0.005, 0.05, 0.5, 5), (2) ratio of normal and shear forces applied outside the wheel with five levels (i.e., 0, 0.1, 0.2, 0.3, 0.4), and (3) volume fraction ($f$) with five levels (i.e., 0.7, 0.8, 0.9, 1, 1.1). We then created 125 optimization problems (5 × 5 × 5) per reference design. The rim part (outer ring of the wheel) is set to the non-design space, the center area is set to a fixed boundary condition, and the spoke part is set to the design space. The maximum iteration of optimization was set to 100 as the termination criterion. Generating one design takes 17 s on average by using a computer with AMD 3990X 64-Core 2.90 GHz CPU and 128 GB RAM.

For the reference design, 658 real wheel images were obtained by web crawling. These images were then preprocessed into a 128 × 128 binary matrix represented by $\mathbf{x}_r$ in Eq. (1). We solved 125 optimization problems for 658 reference designs and obtained 82,250 topology optimization designs (658 × 125 = 82,250). Some of the generated designs have similar topologies and shapes. We calculated the pixelwise L1 distance for all designs and removed the designs that are within the threshold of $10^3$. A total of 16,678 designs are obtained with different topologies and shapes.

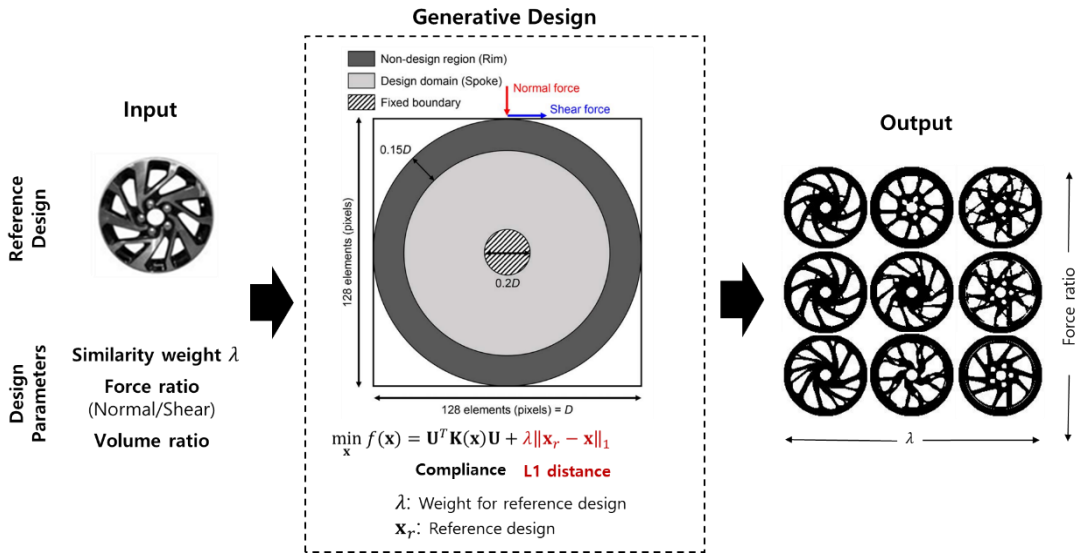

Figure 2. 2D wheel generation by generative design (Oh et al., 2019)

## 4.2. Dimensionality Reduction (Stage 2)

Stage 2 is the process of reducing the dimensions of the 2D wheel design created in Stage 1 using a convolutional autoencoder. Data augmentation was performed to improve the performance, and the latent space was analyzed to verify the feasibility of the trained model.

The autoencoder in this study is used for two purposes: First, it is used for the DOE in latent space. DOE in its original dimension (128 x 128) fails to sample designs that can represent the data distribution. Sampling in latent space can sample a variety of designs that can be representative of the features of 2D geometry. The detailed experimental result is described in Section 4.3. Second, the autoencoder is used for transfer learning. We used the encoder of the pretrained autoencoder to overcome the data shortage of CAE simulations. The effects of improving predictive performance by transfer learning are described in Section 5.2.

### 4.2.1. Data

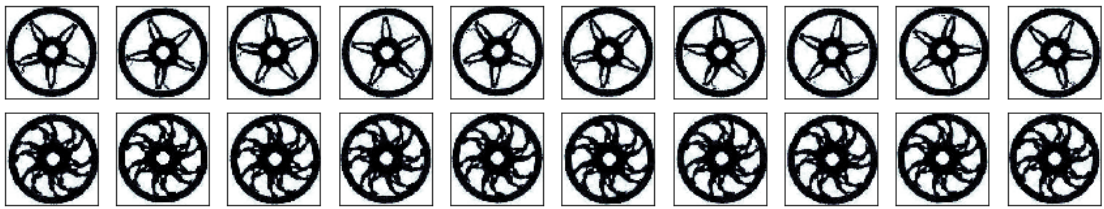

Figure 3. Examples of augmented 2D wheels

Because the wheel is a rotating object, the rotated wheel should not be recognized as a different wheel. Therefore, the data were augmented by rotating randomly in the 360° range. The number of 2D wheel design data increases from 16,678 to 166,812 (approximately 10 times larger) through this augmentation. Figure 3 shows an example of rotating the wheel at 10 random angles. This data augmentation resulted in an increase of the training data size; hence, the deep learning performance improved. The data augmentation effect is experimented in Appendix A.

### 4.2.2. Training

The convolutional autoencoder adds convolutional layers to the autoencoder architecture and performs well in the dimensionality reduction of images (Masci et al., 2011). The architecture used is shown in Figure 4.

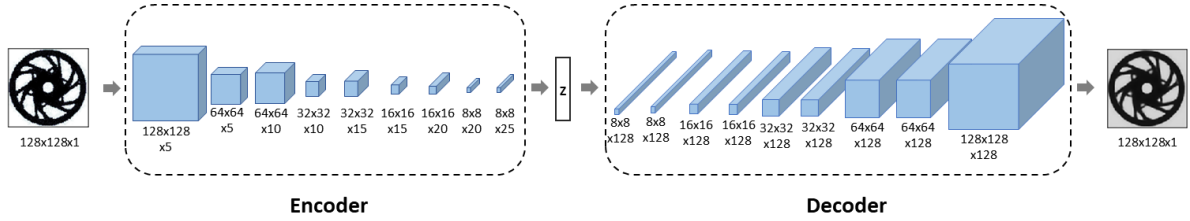

**Figure 4. Architecture of convolutional autoencoder**

When the 2D wheel image measuring 128 × 128 as input passed through the encoder part, the dimension was mapped to 128 dimensions in latent space (**z**); meanwhile, when the 128 dimensioned values were passed through the decoder again, the 2D wheel image that was the input was reconstructed. If the reduced dimension (128) can reconstruct the original dimensions (128 × 128), it demonstrates that the 128 dimensions of **z** well extracted the important features of the input.

An autoencoder is a model that minimizes the difference values of pixels between an input image and an output image. The autoencoder loss function can be expressed as the mean squared error (MSE) as follows:

$$\text{MSE} = \frac{1}{n}\sum_{i=1}^{n}(x_i - \hat{x}_i)^2, \qquad (2)$$

where $x_i$ is the i-th input data, $\hat{x}_i$ is the output from the autoencoder, and $n$ is the number of input data. In the convolutional autoencoder architecture, the encoder is composed of five convolutional layers and four maxpooling layers, and the decoder is composed of five convolutional layers, four upsampling layers, and a 50% dropout; therefore, it becomes a 128 × 128 image again. A rectified linear unit (ReLU) was used for the activation function of each convolution layer. The Adam optimizer was used with a learning rate of 0.00008, batch size of 64, and epoch of 100. A total of 133,449 data points (80%) were used as the training set, whereas 33,363 data points (20%) were used as the validation set. Figure 5 shows the learning results. The model converged well in both the training and validation sets.

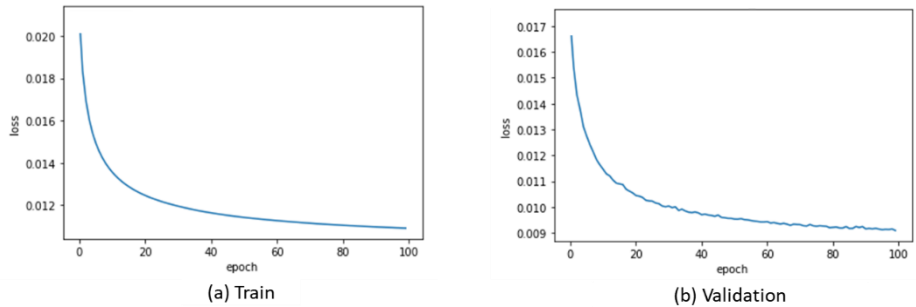

**Figure 5. Reconstruction error of convolutional autoencoder**

### 4.2.3 Testing

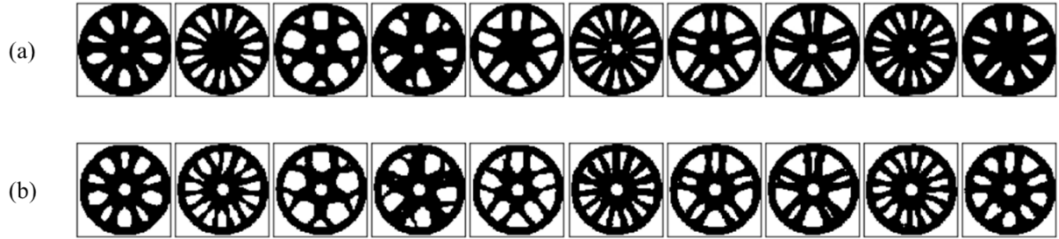

Figure 6. (a) original image (b) reconstructed wheel image in test set

To visually verify whether the model has learned the features well, we reconstructed 96 real wheel images of the manufacturer that were not used for the training and validation. As shown in Figure 6, it was confirmed that the wheels in the test set were reconstructed similarly to the input. Therefore, it can be concluded that the latent space well represented the wheel-shaped features. In addition, in the training set, holes were uniformly drilled in the center of the wheel. Therefore, even if no holes exist in the test data (i.e., the rightmost figure in Figure 6), a hole of the same size is created in the reconstructed wheel. Hence, it can be concluded that the model learned that the center of the wheel is always drilled.

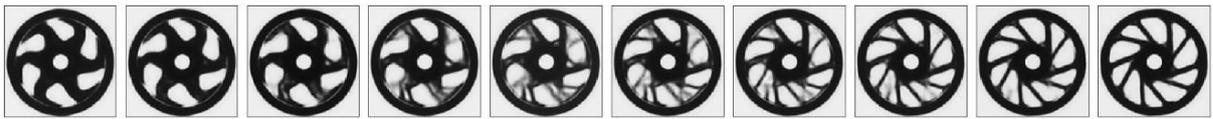

Figure 7. Interpolation result in z latent space

Interpolation is a method to verify whether the deep learning generative model has been well trained for the latent space. When the model only memorizes the input wheel to reconstruct the input wheel, the interpolated wheel image in the latent space does not naturally deform, and the shape of the wheel suddenly changes stepwise. After encoding the two wheels in the training set to two vectors in the latent space, these vectors were divided by the same distance, and 10 vectors were extracted from the latent space. The results of decoding the 10 vectors are shown in Figure 7. Two wheels located on the leftmost and rightmost sides appeared completely different, but the wheels reconstructed from the 10 vectors in latent space changed gradually while maintaining the features of both sides. Therefore, it can be concluded that the convolutional autoencoder has learned the latent space continuously.

## 4.3. DOE in latent space (Stage 3)

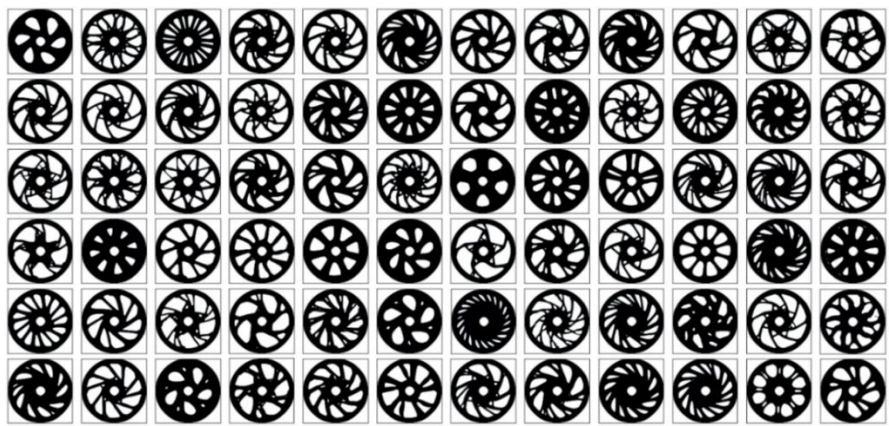

Figure 8. Sampled 2D wheel designs from latent space

LHS from a normal distribution was used to conduct the DOE in the latent space, and the LHSnorm function in MATLAB (Mathworks, 2020) was used. When training data were encoded in the latent space, the data were not distributed uniformly but exhibited a normal distribution. We discovered that a good wheel-shape image can be sampled when we used the LHSnorm and not the LHS. After sampling 3000 vectors from a 128-dimensional multivariate normal distribution with the mean and covariance of the training data in the latent space, the training wheels that were closest to the sampled vectors were selected. The reason for using the nearest training wheel design without using the decoded image was that the main purpose of the convolutional autoencoder was to reduce dimensions, not to generate new images. Finally, we selected 1300 designs after filtering to verify the similarity. Figure 8 shows examples of the sampled wheel design.

We compared the DOE in the original space (128 × 128) to that in the latent space (our method). We recalculated the L1 distance between samples in latent space for the designs sampled in the two methods. The sampling is performed well because the samples are widely distributed in the latent space. In this case, the L1 distance value increases. The designs sampled in the original space have an average distance value of 0.0225, and the designs sampled in the latent space have 0.5693. This finding shows that the DOE in the original space cannot represent the 2D wheel design space well.

### 4.4. 3D CAD Automation (Stage 4)

The process of Stage 4, i.e., creating a 3D CAD model from a 2D image, was performed in three steps, as shown in Figure 9. The first step was image processing, which uses antialiasing to create smooth and sharp edges of the 2D wheel design samples. The second step was data processing by grouping neighboring points to create splines and centering them. The third step was automatic 3D CAD generation using the Python API of Autodesk Fusion360 (Autodesk, 2020b).

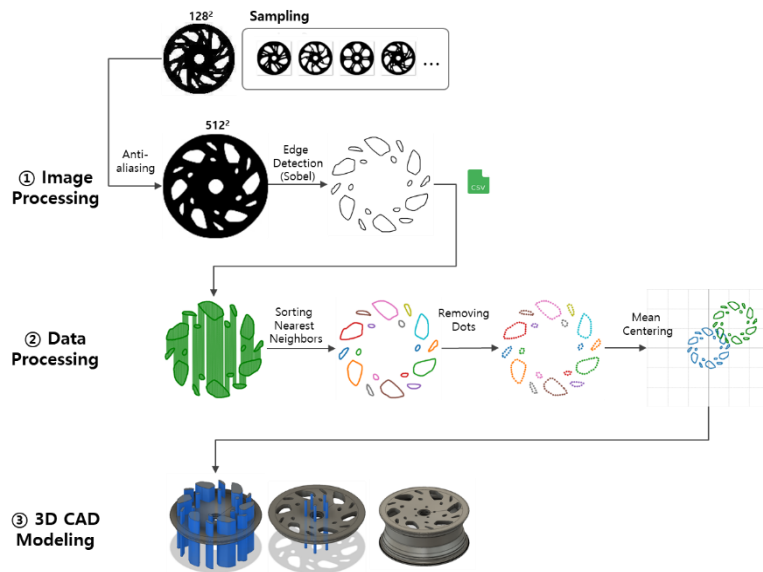

**Figure 9. Three steps of Stage 4**

### 4.4.1. Image Processing

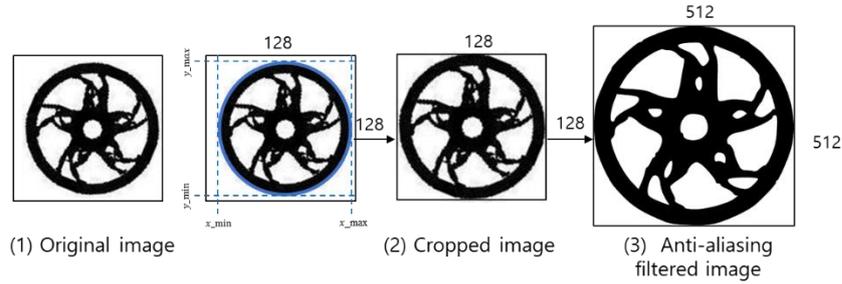

Figure 10. Antialiasing filtering

The original image created in Stage 3 contained a margin at the edge. Because the position of the wheel must be the same in all images, the margin must first be removed. Therefore, we detected the edge, obtained the maximum and minimum values of x and y from the detected data, and cropped the image in the range of $[x_{min} \sim x_{max}, y_{min} \sim y_{max}]$ of the original image. Subsequently, we obtained the image with no margin, as shown in Figure 10.

A low-pixel image appeared as a stepped line owing to the large size of the square pixel; this phenomenon is known as aliasing. In our study, the original image comprised 128 × 128 pixels; therefore, we applied antialiasing (AA) (Catmull, 1978) to all the images, which reduced the antialiasing by smoothing. AA converts a pixel graphic file to a vector graphic file; therefore, we converted the PNG file into the scalable vector graphics format and then converted it back to a PNG file to obtain an antialiased image. In the AA process, a higher pixel image was attainable because more pixels were added to smooth the aliasing portion. Figure 10 shows an example of obtaining a 512 × 512 high pixel image by applying AA to a 128 × 128 wheel image.

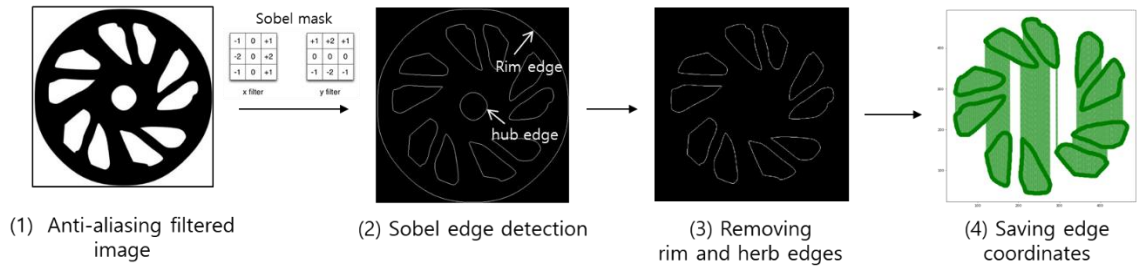

Figure 11. Edge detection

Figure 11 shows the image processing sequence for AA processed images. Edge detection involves obtaining the boundaries of objects in an image, where the boundary is an area with a large change rate in brightness. The change rate in brightness is called the gradient; after calculating the gradient in the image, the edge can be determined according to the threshold value. The first derivative, which is an approximation of the difference between adjacent pixels, was performed in the horizontal and vertical directions. The gradient magnitude $\nabla G$ is calculated as follows:

$$\nabla G = \sqrt{G_x^2 + G_y^2}, \quad (3)$$

where $G_x = f(x+1, y) - f(x, y)$
$G_y = f(x, y+1) - f(x, y)$

The horizontal difference is $G_x$, the vertical difference is $G_y$, and the brightness function is $f$.

A representative method of edge extraction using the first derivative is the Sobel operator (Kanopoulos et al., 1988). The Sobel operator can extract diagonal edges well; therefore, it is well suited for edge detection in 2D wheel designs. After extracting the entire edge of the image, the edges of the rim and hub were removed to use only the edges of the spoke areas. Subsequently, the coordinates of the final edges were saved in a .csv file.

### 4.4.2. Data Processing

Data processing was performed to connect the spoke edges with splines for CAD modeling automation. The data are the coordinates of the 2D pixels, which we call coordinates points.

In this study, we first organized the points based on distance and group adjacent points. Figure 12 shows the sorting and grouping process. We calculated the Euclidean distance between one fixed point and another to obtain the shortest distance. This allows the points to be organized by the shortest distance. If the point closest to the fixed point is greater than or equal to the threshold value, it is regarded as a different group. The detailed process is described in Appendix B.

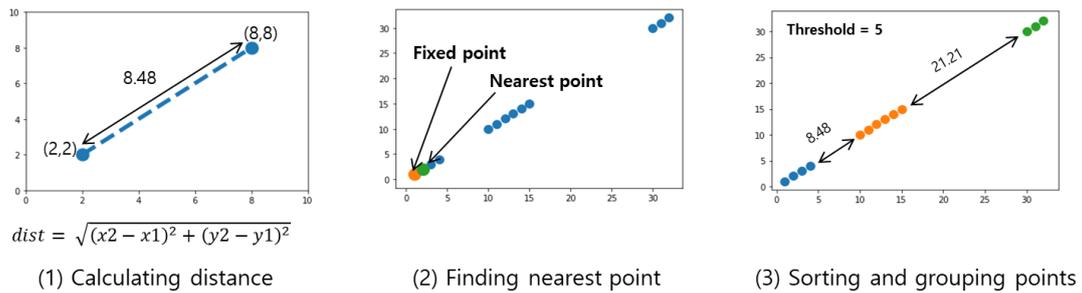

**Figure 12. Algorithm for sorting and grouping points**

If all the points in each group are used for generating the spline curve, the curve will be tortuous and not smooth. This can misrepresent the original shape and cause errors in the CAD modeling. Therefore, the spline curve of the spoke should be designed using the correct number of points after filtering unnecessary points. This process is shown in Figure 13.

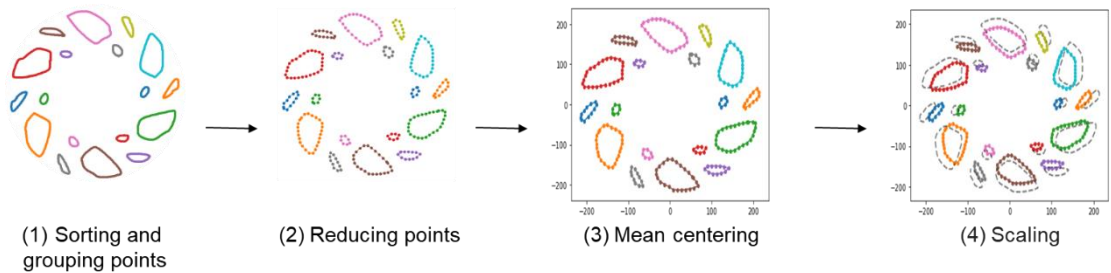

**Figure 13. Data processing**

In this study, we determined the deletion rate according to the number of points in each group. The group comprising more than 20 points and less than 100 points was reduced to 1/6, whereas that comprising 100 points or more was reduced to 1/12. We did not reduce points in a group comprising less than 20 points. However, if the number of points is 3 or less, it is considered as noise and the group is deleted. Finally, all coordinates were moved to the center of the origin (i.e., mean centering). Subsequently, to design a spoke that fit an 18-inch wheel, each point group was multiplied by a scalar, 0.97.

### 4.4.3. 3D CAD Modeling

3D modeling requires not only the disk-view shape, but also the cross-sectional shape of the spokes and rim. The aim of this study is to create wheels with the same cross section and diverse spoke shapes. Typically, in the development of road wheels, the rim cross-section has a limited degree of freedom owing to the packaging constraints of the parts inside the rim. Therefore, we selected the 18-inch representative wheel model with a rim width of $7.5j$, which is a flagship vehicle. $j$ represents the size of the rim and is indicated in inches. From the selected CAD model, the cross sections of the spoke and rim were extracted, and the coordinates of each point were stored in a .csv file. The stored points were used to automatically design cross-sectional shapes through lines and spline curves in 3D modeling, as shown in Figure 14.

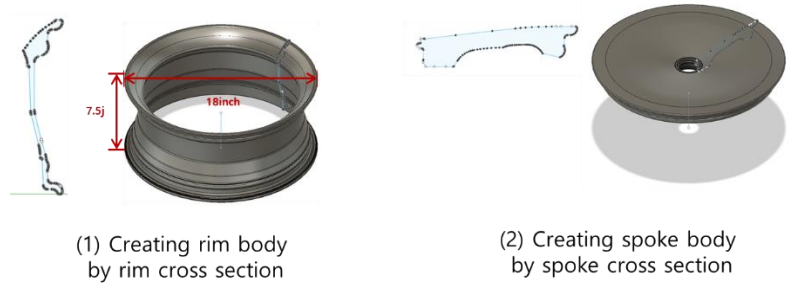

**Figure 14. 3D modeling using selected cross-sections of rim and spokes**

Figure 15 shows the overall process of 3D CAD modeling using the 2D information of disk-view spoke shape and the cross section obtained in previous steps. This process was fully automated using the Python API of Autodesk's Fusion360. The 2D information was obtained from the .csv file and loaded sequentially in the order specified.

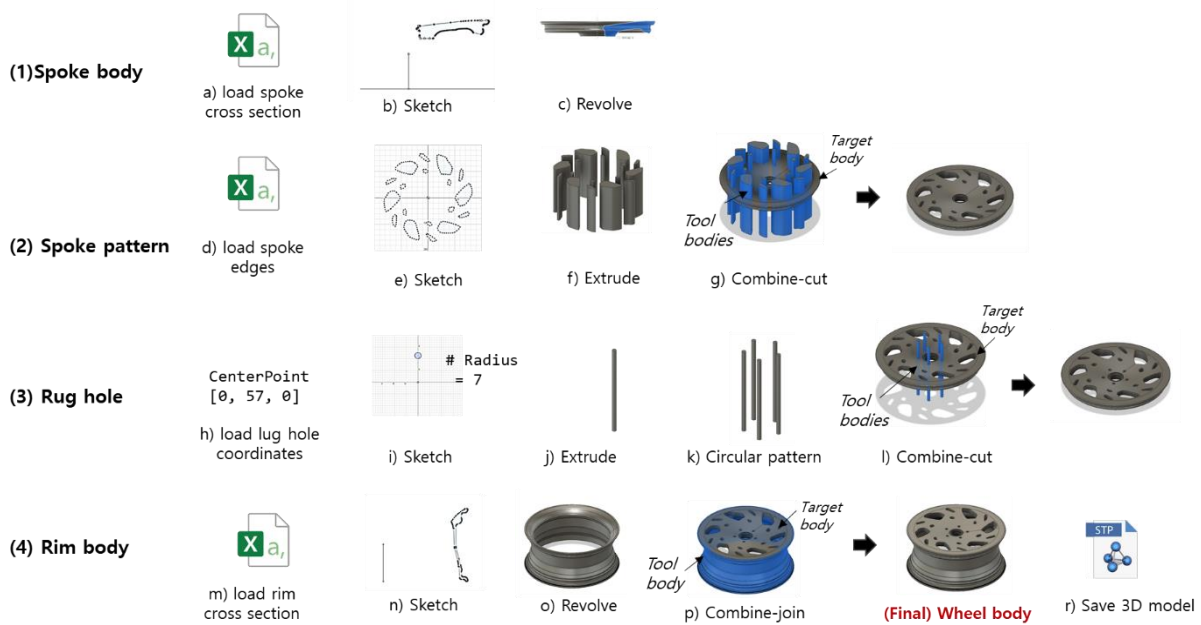

**Figure 15. Process of the automatic generation of 3D wheel CAD using given 2D information**

The detailed processes are as follows:

First, the cross-section of the spoke was sketched by loading the coordinates and connecting the points by lines and splines. When the cross-section was revolved, a spoke body was created.

Second, the disk-view was sketched by loading the coordinates of the spoke edges and connecting the points of each group by a spline. The bodies of the spoke shape were created when sections of the spoke shape were extruded. These spoke-shaped bodies were designated as tool bodies, whereas the spoke body was designated as the target body. Subsequently, a combine-cut was performed to remove the intersection of the two bodies. Consequently, the tool bodies disappeared, and the target body without a cross section remained.

Third, after defining the reference lug hole center coordinates and radius, the circular cross section was sketched and extruded to create a cylindrical body. Subsequently, the same circular column bodies were generated at 72° intervals using the reference body. The five cylindrical bodies created were designated as tool bodies, whereas the spoke body was designated as the target body. A combine-cut was performed to remove the intersection between the bodies, thereby creating lug holes in the spoke body.

Fourth, the cross-section of the rim was sketched by loading the coordinates and connecting the points by the lines and splines. The sketch was then revolved to create a rim body. A combine-join was performed to combine the rim body and spoke body, and then a final wheel body was created. The wheel body was saved in .stp format, which is a universal 3D CAD file format.

Figure 16 shows examples of automatically generated 3D CAD models.

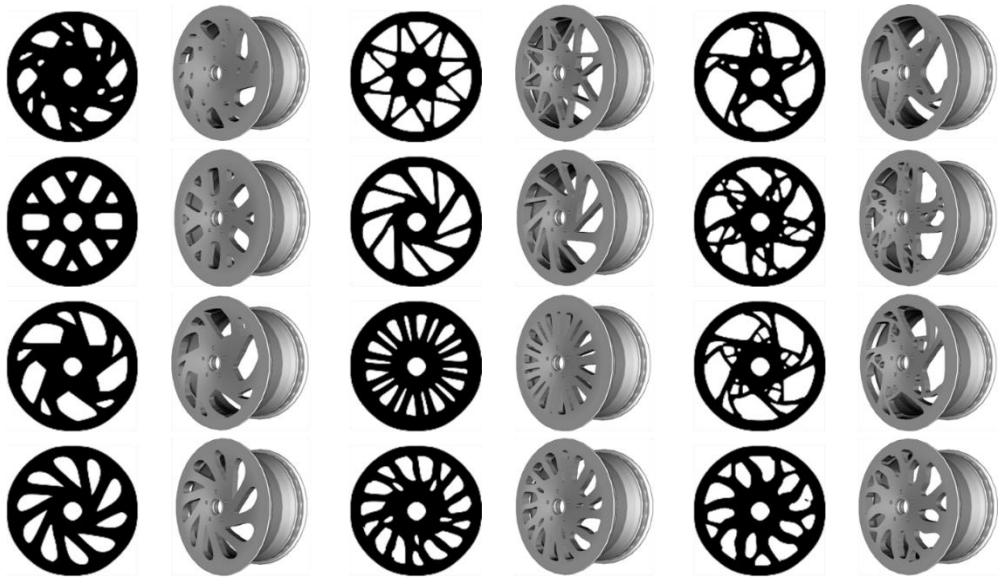

**Figure 16. Auto-generated 3D wheel CAD models**

## 5. Design Evaluation (Stages 5 to 7)

### 5.1. CAE Automation (Stage 5)

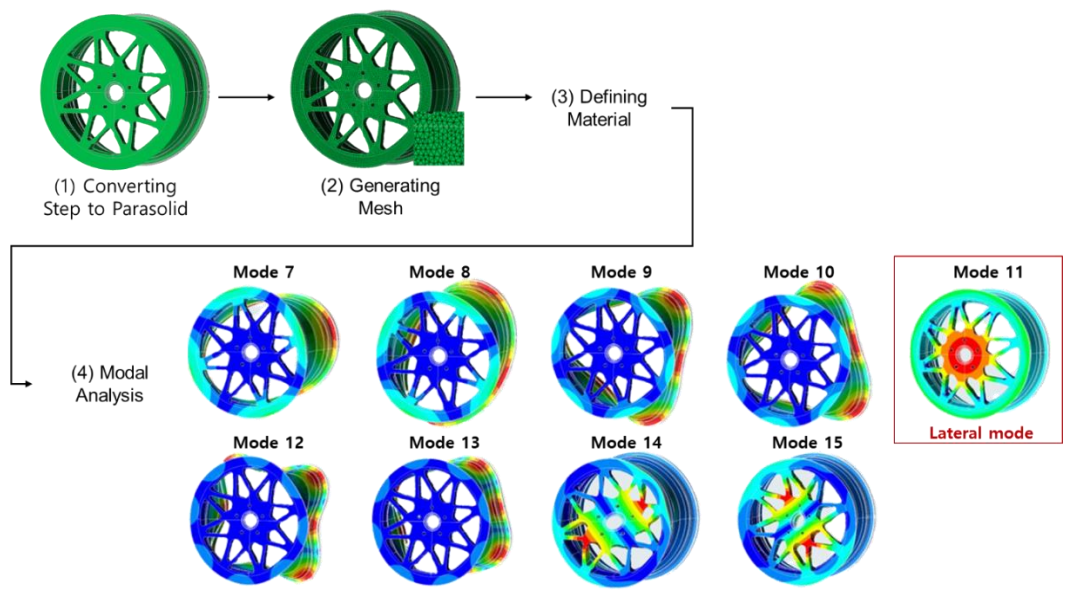

**Figure 17. Example of modal analysis and results**

In this study, modal analysis with free-free mode was conducted to analyze the engineering performance of the wheel. We obtained the natural frequencies and mode shapes through CAE simulation. In particular, the natural frequency is proportional to the stiffness of the structure and inversely proportional to mass as follows:

$$f = \frac{1}{2\pi}\sqrt{\frac{k}{m}}, \tag{4}$$

where $f$ is the natural frequency, $k$ the stiffness, and $m$ the mass. Therefore, when designing a wheel, manufacturers consider a lower bound of stiffness for each mode as a design constraint, based on the correlation between the stiffness for each mode and road noise.

We conducted a free-free modal analysis. The unconstrained 3D model contained six rigid body modes with zero frequency, three translation modes in the x- and y-axis directions and three rotation modes for the three axes. Beginning from the seventh mode, a nonzero frequency appeared. Figure 17 shows the mode shapes of the wheel. Each mode has the following meanings. Modes 7 and 8 indicate the rim mode 1. Modes 9 and 10 indicate the rim mode 2. Mode 11 indicates the spoke lateral mode. Modes 12 and 13 indicate the rim mode 3. Modes 14 and 15 indicate the spoke bending mode.

The frequency for the lateral mode (mode 11) was selected in this study because we intended to evaluate the engineering performance according to the shape of the spoke. In addition, after the mass has been obtained from the 3D modeling, the stiffness can be calculated through the natural frequency and mass using Eq. (4).

For CAE automation, the macro function in Altair's Simlab (Altair, 2020) was used. The modal analysis process is shown in Figure 17. First, the 3D CAD model was imported; next, an FEM mesh (second-order tetrahedral mesh) was automatically generated. The size of the mesh should not be larger than 6 mm. We used the material properties of the reference aluminum wheel of Hyundai: Young's modulus is 73,500, Poisson's ratio is 0.33, density is 2.692e−09, and shear modulus is 0.001. The CAE automation yielded 1,006 output files. The input 2D wheel image, frequency (mode 11), and mass through the automatic parsing program were stored in pairs as training data for deep learning.

## 5.2. Transfer Learning (Stage 6)

In Stage 6, deep learning was developed, in which the results of modal analysis and the mass of the 3D CAD model were predicted using only 2D design. Data augmentation and data scaling were performed as data preprocessing; furthermore, transfer learning and ensemble techniques, in which the pretrained convolutional autoencoder and DNN were combined, were applied to solve the problem of insufficient data.

### 5.2.1. Data

Data augmentation was performed to avoid overfitting. We augmented the 1006 wheel designs that we sampled in Stage 3 10 times, rotating each of them by 72° and flipping them left and right. Consequently, 10,060 samples were used for training. The output values, mode frequency, and mass were equivalent because imposing the left and right flips did not affect the modal analysis and mass result. The 10 augmented designs are used as different input data for deep learning, but the output values (labels) are the same.

Furthermore, 80% of the 10,060 data was used as the training set, and the other 20% was used for validation. Although a new design that does not exist in the market because it was created with a generative design was used for training and validation, 96 wheel images was used for the test set, which were sold by the manufacturer. This was to confirm whether the trained model can predict the actual data of the manufacturer. The frequency and mass values of the test set were obtained through Stages 4 and 5, respectively, identical to the training data.

We used min–max scaling, also known as normalization (Al Shalabi and Shaaban, 2006). This method was adjusted to a fixed range (regularly 0 to 1) for the output labels. The formula for $y_{scale}$ is as follows:

$$y_{scale} = \frac{y - y_{min}}{y_{max} - y_{min}} \tag{5}$$

### 5.2.2. Training

In this study, to obtain the optimal architecture and assess the degree to which transfer learning and ensemble affects performance, we trained four models that corresponded to Table 1.

**Table 1 Four deep learning architectures**

| Models | Description |
|---|---|
| CNN | A model that uses only a CNN regressor and not transfer learning (added seven fully connected floors, four max pooling layers, and five convolution layers, i.e., the same structure as the encoder part of the convolutional autoencoder). |
| TL_VGG16 | A model that uses transfer learning through a pretrained VGG16 (using ImageNet dataset). |
| TL_CAE | A model that uses transfer learning through the pretrained convolutional autoencoder model (see Figure 18). |
| TL_CAE_Ensemble | A model that applies ensemble in TL_CAE and trains using the averaging technique for nine frequency models and five mass models (see Figure 19). |

First, the CNN was used as a baseline model.

Second, TL_VGG16 is a model that uses a trained VGG16 (Simonyan and Zisserman, 2014) as transfer learning. Transfer learning transfers the pretrained deep learning model from the domain where data are abundant, which is a method to train a domain that lacks data; moreover, it is one of the most used methodologies in deep learning because it can accomplish a high accuracy despite insufficient data. (Rawat and Wang, 2017)

Third, TL_CAE is a representative model used in this study, which transfers the weight and architecture of the encoder of the convolutional autoencoder model pretrained using 166,812 data in Stage 2, adds fully connected layers as a regressor, and performs fine-tuning using 10,060 data from the modal analysis result. Figure 18 shows a visualization of the deep learning architecture used in this study. Although we performed an augmentation (as mentioned in Section 8.1), the number of modal analysis results (label) was 1,006, which was small but expected to improve the performance using the encoder pretrained using 166,812 data (a large amount).

The newly added regressor part comprised seven fully connected layers. In the fully connected layers, all nodes of a layer are connected to all other nodes of subsequent layers to make decisions for regression and help extract the global relationship between the features. The number of fully connected layers was chosen through random search (trial and error). TL_CAE with seven FC layers improved the RMSE and MAPE by around 12% and 10% compared with TL_CAE with one FC layer.

It was trained using the Adam optimizer, in which the learning rate was 0.002, decay rate was 0.001, and batch size was 256. Early stopping, which is a method used to avoid overfitting when training a model, was applied. Early stopping rules stop training when the error on the validation set is higher than the last time it was checked, that is, when the model performance did not improve for the validation dataset.

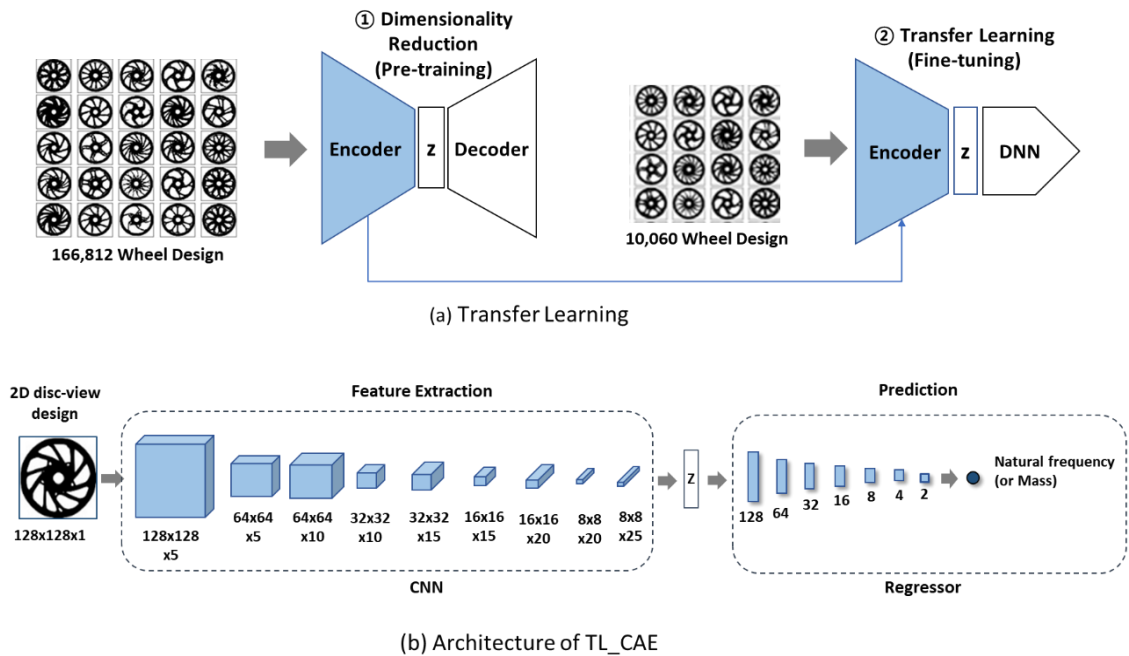

(a) Transfer Learning

(b) Architecture of TL_CAE

**Figure 18. Transfer learning using convolutional autoencoder (TL_CAE)**

Fourth, the final model TL_CAE_Ensemble was built using the ensemble technique, as shown in Figure 19. The ensemble technique offers the advantage of avoiding overfitting in regression problems and reducing bias deviation (Geurt et al., 2006). It is a method of standardizing the prediction results by assembling multiple models into a single model. Our ensemble model uses the mean values of nine frequency prediction results and five mass prediction results. For the frequency prediction model, the training required 15 min on four GPUs (GTX 1080) in parallel.

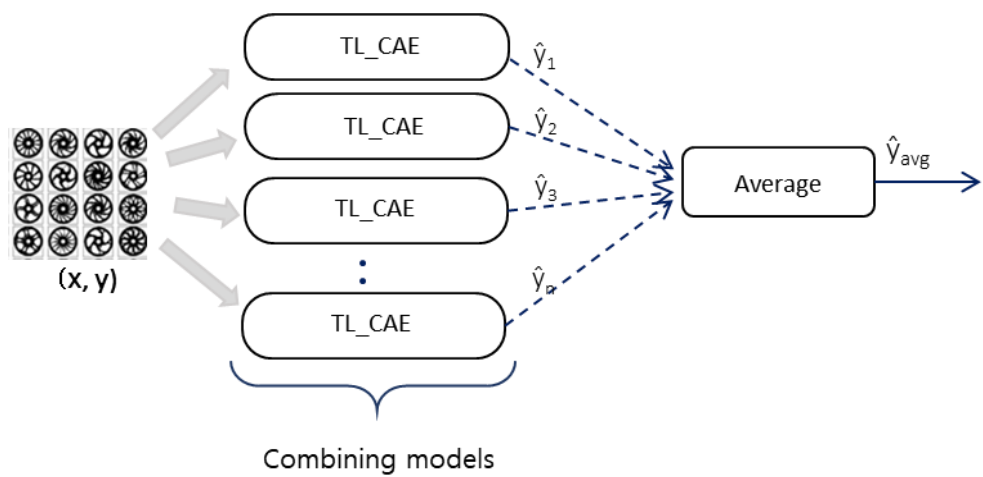

**Figure 19. Ensemble model for TL_CAE (TL_CAE_Ensemble)**

### 5.2.3. Testing

To evaluate the performance of the prediction model, two metrics, the root mean square error (RMSE) and the mean absolute percent error (MAPE) were used. They are expressed as follows:

$$RMSE = \sqrt{\frac{1}{n}\sum_{i=1}^{n}(\hat{y}_i - y_i)^2} \tag{6}$$

$$MAPE = \frac{100}{n}\sum_{i=1}^{n}\left|\frac{\hat{y}_i - y_i}{y_i}\right| \tag{7}$$

$\hat{y}$ is the predictive value, $y$ the ground truth value, and $n$ the number of data points. The performance results of the four trained models are shown in Table 2. TL_CAE indicated improved RMSE and MAPE compared with TL_VGG16. This result confirmed that the proposed convolutional-autoencoder-based transfer learning performed effectively. In addition, the final model, TL_CAE_Ensemble, demonstrated the highest predictive performance in terms of RMSE and MAPE.

The errors for the validation and test sets of the three models are represented as histograms, as shown in Figure 20. The closer the error is to 0, the higher is the accuracy of the model. These results confirmed the effects of transfer learning and the ensembles.

**Table 2 Comparison of frequency and mass prediction results**

*units: Frequency (Hz), Mass (kg)

| Method | Training set | | | | Validation set | | | | Test set | | | |
|---|---|---|---|---|---|---|---|---|---|---|---|---|
| | Frequency | | Mass | | Frequency | | Mass | | Frequency | | Mass | |
| | RMSE | MAPE | RMSE | MAPE | RMSE | MAPE | RMSE | MAPE | RMSE | MAPE | RMSE | MAPE |
| CNN | 13.0 | 1.06 | 0.10 | 0.48 | 16.71 | 1.24 | 0.09 | 0.44 | 20.97 | 3.51 | 0.33 | 1.89 |
| TL_VGG16 | 4.58 | 0.36 | 0.26 | 1.46 | 6.57 | 0.50 | 0.27 | 1.49 | 20.56 | 1.74 | 0.19 | 0.91 |
| TL_CAE | 8.89 | 0.69 | 0.09 | 0.44 | 10.44 | 0.82 | 0.09 | 0.45 | 18.89 | 1.38 | 0.13 | 0.67 |
| TL_CAE_Ensemble | **7.35** | **0.56** | **0.06** | **0.28** | **8.72** | **0.66** | **0.06** | **0.29** | **12.78** | **0.90** | **0.12** | **0.54** |

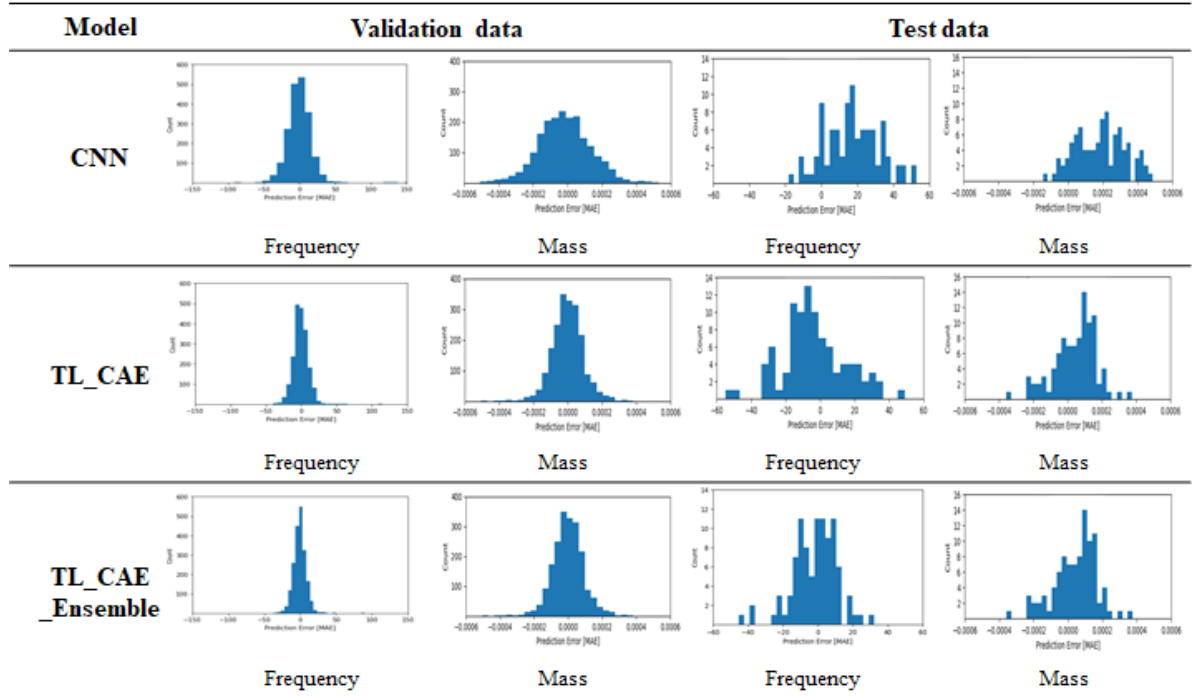

**Figure 20. Comparison of errors**

We can evaluate a large amount of generated wheel concepts or new conceptual designs by calculating the stiffness in terms of the predicted frequency and mass through the proposed deep learning model. Evaluating a wheel concept using a computer with NVIDIA TITAN Xp 4.8 GB GPU takes 0.66 s on average. This model can have many advantages when we use multiple GPUs for parallel computing. Figure 21 shows an example of selecting the wheels in the order of highest stiffness in the validation set. An automaker has its own stiffness standard, and unsatisfactory designs can be eliminated on the basis of this value. Various engineering performance criteria should be evaluated beyond modal analysis, and trade-offs can be found between performance (Oh et al., 2019). On the basis of the evaluation results, the engineer can choose a conceptual design candidate to use in the detailed design phase.

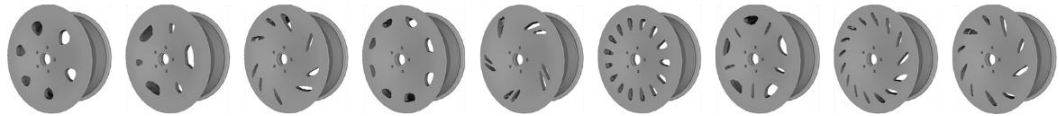

Figure 21. Wheel design candidates in the order of highest stiffness (from left to right)

### 5.3. Visualization and Analysis (Stage 7)

This section describes the visualization and deep learning results from using the proposed framework, and the approach to ensure the reliability of the results.

#### 5.3.1. Feature Visualization

We visualized the latent space of the convolutional autoencoder in Stage 2 to analyze the explainable features of the generated wheel shape. We embedded 16,678 wheel data into the 128 dimensions of the latent space, and then grouped the data by K-means clustering. In the K-means method, the sum of squared errors (SSE) was plotted over the number of clusters. Additionally, the value corresponding to the elbow, which is the part where the SSE reduction rate decreased rapidly, was determined as the number of clusters. Because the shape of the wheel varied significantly, a large number of groups was required. However, we arbitrarily set the number of clusters to 20 because an excessive number would complicate intuitive analyses by CAD/CAE engineers. In Figure 22, for the visualization, the latent space of 128 dimensions was reduced to two dimensions again through T-SNE (Maaten & Hinton, 2008) to display the data, and the colors of the 20 groups are displayed. In addition, the example wheels of each group were selected and shown.

Commercially available road wheels can be classified into the dish, spoke, mesh, fin, and spiral types according to the shape (Napac, 2020); furthermore, these types can be compounded on one wheel. In this study, the spiral type is the most typically generated. This is because shear force was applied to generative design. In addition, in terms of spoke thickness, groups 4, 7, 12, 16, 17, and 19 were relatively thin, whereas groups 6, 11, 13, and 15 were relatively thick. This visualization of latent space facilitates the understanding of the geometric meaning of positions in a latent space.

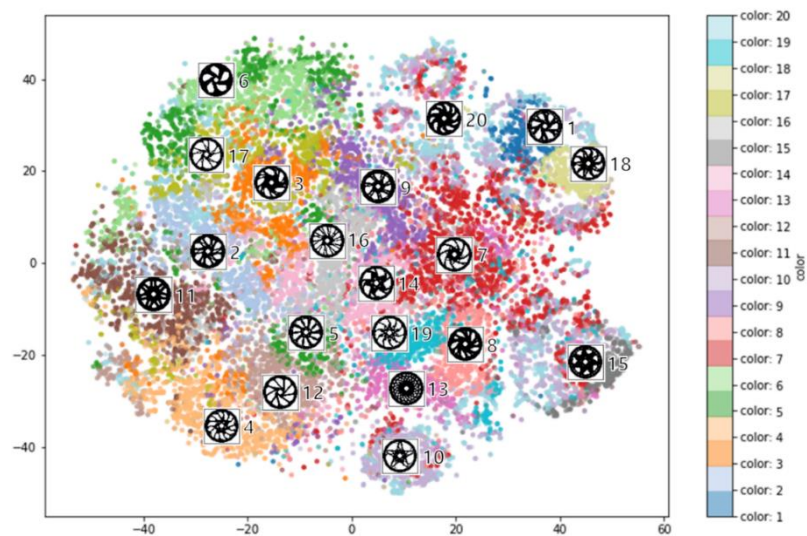

Figure 22. Visualization of latent space using T-SNE

### 5.3.2. Relationship Between Features and Engineering Performance

As shown in Figure 23, visualization was performed to determine if the modal analysis result, which is the engineering performance, can be explained by the features of the wheel shape. We embedded 1006 wheels that were used in the modal analysis in Stage 5 into the latent space and visualized them in a two-dimensional plane using T-SNE. In addition, the magnitude of the natural frequency value of each wheel is displayed in color. The frequency value was categorized into 10 groups using K-means. Higher frequencies are represented in red, whereas lower ones, in blue.

Figure 23 shows that wheels with similar natural frequencies accumulated in the latent space, which represents the wheel shape. It was confirmed that the disk-view shape of the wheel was highly correlated with the lateral mode frequency. As shown in Figure 24, we sampled example wheels from each frequency group; the results show that the thicker the spoke, the more the frequency increases.

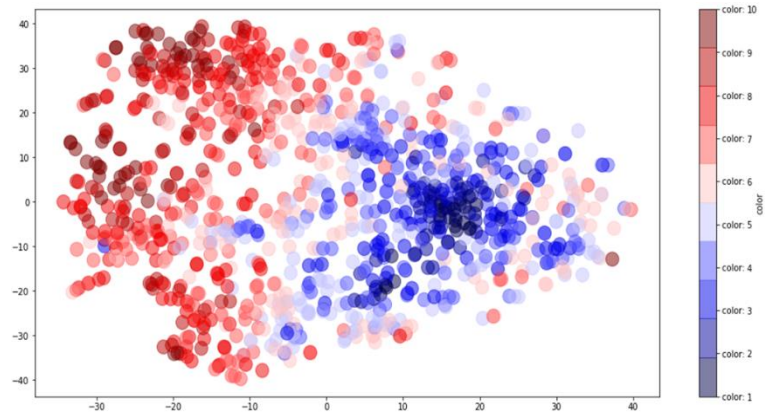

Figure 23. Visualization of latent space with lateral mode frequency

| Group | 1 | 2 | 3 | 4 | 5 | 6 | 7 | 8 | 9 | 10 |
|---|---|---|---|---|---|---|---|---|---|---|
| Frequency | 736-797 | 802-863 | 864-910 | 911-949 | 949-986 | 987-1025 | 1025-1066 | 1068-1111 | 1112-1156 | 1157-1221 |

Figure 24. Wheel design of each lateral mode frequency group

This visualization enables clusters of high-performance designs to be selected and important design features to be analyzed intuitively. Furthermore, it provides a method to visually verify the reliability of CAD/CAE results and gain insights into better performance designs.

### 5.3.3. Grad-CAM

The high predictive performance afforded by AI are difficult to explain. A trained CNN model includes the number of weights and biases. The magnitude of these weights and biases themselves cannot explain which areas of the wheel design affect natural frequencies. Hence, the importance of eXplainable AI (XAI) research is increasing. A representative XAI technology, Grad-CAM (Selvaraju et al., 2017), was applied to the proposed framework, allowing CAD/CAE engineers to make reliable decisions based on deep learning results.

A class activation map (CAM) can be used to visualize important areas of input data (images) that significantly affect the classification output in the CNN model (Zhou et al., 2016). A CAM is obtained by adding global average pooling (GAP), a FC layer, and softmax to the feature map, which is the output of the last convolutional layer. The weight of the FC layer indicates the importance of each feature map to the target class label. After multiplying each feature map by weights and adding them together, we can obtain a CAM that can be visualized using a heat map. The disadvantage of the CAM is that GAP must be added to the CNN architecture and then the model must be trained again. However, Grad-CAM does not require the CNN architecture to be modified. This is because Grad-CAM uses gradients obtained through backpropagation instead of using the weights of the FC layer.

Originally, Grad-CAM was proposed for the classification problem; however, we modified it in our study to adapt to the regression problem. The equation of the regression score $L_{Grad-CAM}$ is as follows:

$$L_{Grad-CAM} = ReLU(\sum_k a_k A^k), \tag{8}$$

where $a_k = \frac{1}{z}\sum_i \sum_j \frac{\partial y}{\partial A_{ij}^k}$

$A_{ij}^k$ indicates the value corresponding to $i$ rows and $j$ columns of the $k$-th feature map, and $a_k$ is the GAP result of the partial derivative of $y$ by $A_{ij}^k$. After linearly combining $a_k$ with the feature map $A^k$, the ReLU activation function was applied to obtain Grad-CAM, which can highlight important areas in the image.

Figure 25 shows the visualization result by applying Grad-CAM to the TL_CAE model. Ten example wheels were selected from the 10 frequency groups shown in Figure 24, and the Grad-CAM for each wheel is displayed as a heatmap in the third row of Figure 25. In the second row of Figure 25, the superimposed image shows the common area of the 2D wheel and the Grad-CAM. The results indicate that an important area (highlighted in red) that affected the frequency value was the central part of the wheel. It appeared that the frequency increased as the center was filled. This is because the largest displacement occurred in the center of the wheel in the lateral mode shape (see Figure 17). Therefore, Grad-CAM confirmed that deep learning results can explain the physics of the lateral mode shape and ensure high reliability predictions.

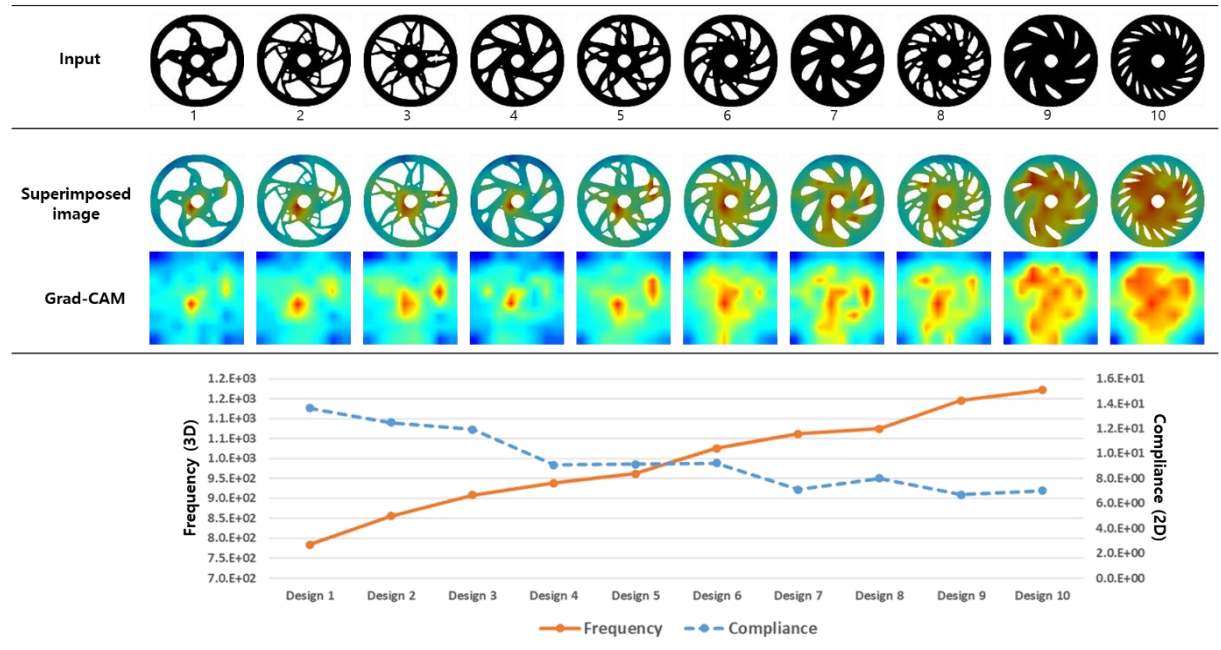

**Figure 25. Result of Grad-CAM for frequency of lateral mode**

# 6. Discussion and Conclusion

## 6.1. Discussion

6.1.1. Generation versus evaluation

The proposed framework can be divided into a generation phase and an evaluation phase, and the engineering design criteria used in each phase have different purposes. In the generation phase, we solve a multiobjective optimization problem by minimizing compliance and maximizing similarity with a reference design. This formulation with reference design data results in creating a large amount of stiff 2D designs that have a real wheel-like shape. The reason for using stiffness optimization for 2D design rather than natural frequency optimization is that this approach has been proven in previous work to be suitable for generating a variety of wheel designs (Oh, et al., 2016). The modal analysis for 3D design was chosen because of the automotive company's requirements. In future work, we plan to apply the natural frequency optimization for both 2D and 3D. However, the design diversity in this case can be reduced because the force conditions (e.g., ratio of normal and shear forces applied outside of the wheel), such as the compliance minimization problem, cannot be varied.

Solving multiobjective design optimization considering compliance and natural frequency can be an alternative (Ahmed et al., 2016). However, the design quality and design diversity in generative design have trade-offs (Chen & Ahmed, 2021). Thus, designers have to sacrifice design diversity and lose the possibility of exploring novel designs if they consider many design requirements. Therefore, we suggest that only important constraints must be considered in the generation phase.

In the evaluation phase, all different engineering performance can be evaluated. The modal analysis used in this study is only one example of engineering performance. Various evaluations, such as stiffness analysis (i.e., rim stiffness and disk stiffness), strength analysis for impact tests, and aerodynamics, are required to develop a new wheel. This evaluation allows us to filter the wheels created during the generation phase.

A reason for generating designs first and then evaluating the various engineering performance is to make the proposed process easy to use in the industry. The evaluation standards for engineering performance and the importance of each performance are always different in accordance with product concepts. Thus, we can choose the design depending on our development purpose if we create a large number of designs first.

6.1.2. 3D generative design through 2D design

The ultimate goal of our future research is performing 3D generative design that do not go through 2D. However, many problems are needed to be solved in creating various real product-like 3D designs directly through 3D topology optimization. Collecting 3D reference design data is inapplicable, and performing topology optimization that resembles 3D reference design is difficult to be converged due to high dimentionality. In other words, directly implementing the 2D generative design formulation, Eq. (1), into the 3D problem is unsuitable. The higher computational cost for 3D topology optimization compared to that of 2D optimization is also a bottleneck for generating large number of conceptual design data sets.

Conventional topology optimization without maximizing similarity to a reference design generates unfamiliar designs that are far from product designs in the market. Figure 26 shows the conventional 3D topology optimization results with compliance minimization. 3D topology optimization results do not have a familiar shape compared with the wheels generated by the proposed method (Figure 16).

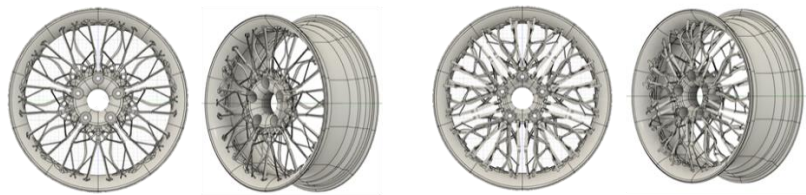

Figure 26. 3D topology optimization examples without using a reference design

6.1.3. Comparison with previous design approach

This work modified conventional design methods to fit the deep learning framework. Compared with the existing methodology, the advantages of the proposed framework are as follows. First, the proposed framework generates a variety of topology designs using topology optimization-based generative design and deep learning methods. Design methods, such as topology optimization and parametric design, are similar. However, the goal of conventional topology optimization is to find the single optimal design rather than exploring diverse designs, similar to the proposed method. Parametric design by changing certain geometry is useless for exploring new topology designs. Second, the proposed framework performs DOE with the design features in the latent space. Section 4.2.3 shows that in the original space, DOE does not effectively represent the 2D wheel design space. Finally, the proposed framework creates a 3D CAD model through a 2D topology design. It can efficiently create a larger number of 3D designs than 3D topology optimization. This framework has the advantage of creating new designs that are similar to existing product designs on the market, whereas traditional 3D topology optimization creates unfamiliar designs that are far from the products on the market, as shown in Figure 26.

However, the proposed method has some limitations. This study has the same drawback as the surrogate modeling study. In particular, building a deep learning model requires collecting a large amount of data, thereby inevitably resulting in a large number of CAE simulations (e.g., 1,006 simulations for this study). However, for companies aiming to quickly evaluate a large number of conceptual designs, such front loading can improve the efficiency of many subsequent product design projects. Performing CAE simulations for each design is extremely inefficient when AI can automatically generate a large amount of conceptual designs. The proposed framework cannot address the sufficient manufacturing constraints, which are important criteria for the design. In the current work, we used density filtering with Heaviside projection to prevent checkerboard patterns and achieved symmetry by resembling a reference design.

## 6.2. Conclusion

This study proposes a deep learning-based CAD/CAE framework for the conceptual design phase. This framework can automatically generate numerous feasible 3D CAD data and immediately evaluate their engineering performance. In the conceptual design stage, industrial designers and engineers can obtain a large number of 3D CAD models by using the proposed framework along with the engineering performance result estimated by AI and discuss suitable conceptual design candidates for the detailed design stage. The proposed deep learning model can predict the CAE results in terms of on 2D disk-view design. Industrial designers can obtain instant feedback regarding the engineering performance of 2D concept sketches.

The framework is proposed for practical use in real product development by integrating various deep learning techniques and the existing CAD/CAE processes. The proposed framework comprised seven stages: (1) 2D generated design, (2) dimensionality reduction, (3) DOE in latent space, (4) CAD automation, (5) CAE automation, (6) transfer learning, and (7) visualization and analysis. The proposed framework was demonstrated through a case study pertaining to road wheel design.

This study contributes to the generation and evaluation of conceptual designs, and its summary is as follows. First, for the generation of designs (stages 1 to 4), the proposed framework provides a solution to the difficulty in obtaining 3D CAD training data for deep learning. 3D wheel CAD data are automatically generated in large quantities by integrating 2D generative design and 3D automation techniques. Second, for the evaluation of designs (stages 5 to 7), the proposed framework provides a deep learning-based surrogate model to evaluate conceptual wheel designs and explain the relationship between geometry and engineering performance, thereby enabling industrial designers and engineers to make reliable decisions.

The future research plan is as follows. First, deep learning will be applied using 3D data as input through the preprocessing of voxels and point clouds. Second, deep learning will be applied to various CAE simulations to predict nonlinear and dynamic analysis results for the design evaluation phase. Third, we will study a generative design method that considers aesthetics on the basis of customer's choice data (Kang et al., 2019). Fourth, manufacturing constraints will be considered. We plan to consider length scales for 3D design evalution. Fifth, the out-of-plane stiffness should be considered in 2D design. Finally, we plan to develop a 3D generative design method without using 2D images.

## Acknowledgements

This work was supported by Hyundai Motor Company and the National Research Foundation of Korea (NRF) grants funded by the Korean government [grant numbers 2017R1C1B2005266, 2018R1A5A7025409]. The authors would like to thank Hyundai Motor Company's Jiun Lee, Sangmin Lee, Min Kyoo Kang, ChangGon Kim, and ChulWoo Jung for their valuable feedback and ideas on our research. We would also like to thank Altair Korea's Jeongsun Lee and Seung-hoon Lee for their help in automating the CAE process.

## Declaration of interests

The authors declare that they have no known competing financial interests or personal relationships that could have appeared to influence the work reported herein.

## Replication of results

The code used in this paper cannot be disclosed as it is the property of Hyundai Motor Comapny that funded this study.

# References

Ahmed, F., Deb, K., & Bhattacharya, B. (2016). Structural topology optimization using multi-objective genetic algorithm with constructive solid geometry representation. Applied Soft Computing, 39, 240–250.

Al Shalabi, L., & Shaaban, Z. (2006, May). Normalization as a preprocessing engine for data mining and the approach of preference matrix. In 2006 International conference on dependability of computer systems (pp. 207-214). IEEE.

Altair. (2019). SimLab. Retrieved from https://www.altair.com/simlab/

Andreassen, E., Clausen, A., Schevenels, M., Lazarov, B. S., & Sigmund, O. (2011). Efficient topology optimization in MATLAB using 88 lines of code. Structural and Multidisciplinary Optimization, 43(1), 1-16.

Autodesk. (2020a). Generative design. Retrieved from https://www.autodesk.com/solutions/generative-design/

Autodesk. (2020b). Retrieved from https://www.autodesk.com/products/fusion-360/

Bendsoe, M. P., & Kikuchi, N. (1988). Generating optimal topologies in structural design using a homogenization method.

Bourdin B (2001) Filters in topology optimization. Int J Numer Methods Eng 50(9): 2143–2158

Bruns TE, Tortorelli DA (2001) Topology optimization of nonlinear elastic structures and compliant mechanisms. Comput Methods Appl Mech Eng 190(26–27): 3443–3459

Burnap, A., Liu, Y., Pan, Y., Lee, H., Gonzalez, R., & Papalambros, P. Y. (2016, August). Estimating and exploring the product form design space using deep generative models. In ASME 2016.

Chen, W., & Ahmed, F. (2021). PaDGAN: Learning to Generate High-quality Novel Designs. Journal of Mechanical Design, 143(3).

Cunningham, J. D., Simpson, T. W., & Tucker, C. S. (2019). An investigation of surrogate models for efficient performance-based decoding of 3D point clouds. Journal of Mechanical Design, 141(12).

Du, X., Sun, C., Zheng, Y., Feng, X., & Li, N. (2020a). Evaluation of vehicle vibration comfort using deep learning. Measurement, 108634.

Du, X., Xu, H., & Zhu, F. (2020b). Understanding the effect of hyperparameter optimization on machine learning models for structure design problems. arXiv preprint arXiv: 2007.04431.

Edwin Catmull. (1978). "A hidden-surface algorithm with anti-aliasing." In Proceedings of the 5th annual conference on Computer graphics and interactive techniques (SIGGRAPH '78).

Feng, Y., Feng, Y., You, H., Zhao, X., & Gao, Y. (2019, July). MeshNet: mesh neural network for 3D shape representation. In Proceedings of the AAAI Conference on Artificial Intelligence (Vol. 33, pp. 8279-8286).

Guest, J. K., Prévost, J. H., & Belytschko, T. (2004). Achieving minimum length scale in topology optimization using nodal design variables and projection functions. International journal for numerical methods in engineering, 61(2), 238–254.

Geurts, P., Ernst, D., & Wehenkel, L. (2006). Extremely randomized trees. Machine learning, 63(1), 3-42.

Guo, X., Li, W., & Iorio, F. (2016, August). Convolutional neural networks for steady flow approximation. In Proceedings of the 22nd ACM SIGKDD International Conference on Knowledge Discovery and Data Mining (pp. 481-490). ACM

Hinton, G. E., & Salakhutdinov, R. R. (2006). Reducing the dimensionality of data with neural networks. science, 313(5786), 504-507.

International Design Engineering Technical Conferences and Computers and Information in Engineering Conference (pp. V02AT03A013-V02AT03A013). American Society of Mechanical Engineers

Jang, S., Yoo, S., & Kang, N. (2021). Generative Design by Reinforcement Learning: Enhancing the Diversity of Topology Optimization Designs. arXiv preprint arXiv:2008.07119.

Kallioras, N. A., & Lagaros, N. D. (2020). DzAIℕ: Deep learning based generative design. Procedia Manufacturing, 44, 591-598.

Kanezaki, A., Matsushita, Y., & Nishida, Y. (2018). Rotationnet: Joint object categorization and pose estimation using multiviews from unsupervised viewpoints. In Proceedings of the IEEE Conference on Computer Vision and Pattern Recognition (pp. 5010-5019).

Kang, N., Ren, Y., Feinberg, F., & Papalambros, P. (2019). Form+ function: Optimizing aesthetic product design via adaptive, geometrized preference elicitation. arXiv preprint arXiv:1912.05047.

Kanopoulos, N., Vasanthavada, N., & Baker, R. L. (1988). Design of an image edge detection filter using the Sobel operator. IEEE Journal of solid-state circuits, 23(2), 358-367.

Khadilkar, A., Wang, J., & Rai, R. (2019). Deep learning–based stress prediction for bottom-up SLA 3D printing process. The International Journal of Advanced Manufacturing Technology, 102(5-8), 2555-2569

Krish, S. (2011). A practical generative design method. Computer-Aided Design, 43(1), 88-100.

Krizhevsky, A., Sutskever, I., & Hinton, G. E. (2012). Imagenet classification with deep convolutional neural networks. In Advances in neural information processing systems (pp. 1097-1105).

Kunakote, T., & Bureerat, S. (2011). Multi-objective topology optimization using evolutionary algorithms. Engineering Optimization, 43(5), 541-557.

LeCun, Y., Bengio, Y., & Hinton, G. (2015). Deep learning. Nature, 521(7553), 436-444.

Maaten, L. V. D., & Hinton, G. (2008). Visualizing data using t-SNE. Journal of Machine Learning Research, 9, 2579-2605.

Masci, J., Meier, U., Cireşan, D., & Schmidhuber, J. (2011, June). Stacked convolutional auto-encoders for hierarchical feature extraction. In *International Conference on Artificial Neural Networks* (pp. 52-59). Springer, Berlin, Heidelberg.

Matejka, J., Glueck, M., Bradner, E., Hashemi, A., Grossman, T., & Fitzmaurice, G. (2018, April). Dream lens: Exploration and visualization of large-scale generative design datasets. In Proceedings of the 2018 CHI Conference on Human Factors in Computing Systems (pp. 1-12).

Mathworks [Computer software]. (2020). Retrieved from https://mathworks.com/.

Maturana, D., & Scherer, S. (2015, September). Voxnet: A 3d convolutional neural network for real-time object recognition. In 2015 IEEE/RSJ International Conference on Intelligent Robots and Systems (IROS) (pp. 922-928). IEEE.

Napac. (n.d.). Light Alloy Wheel Categorization by Design. Retrieved from https://www.napac.jp/cms/en/wheel-words/wheel-design-types.

Nie, Z., Jiang, H., & Kara, L. B. (2020). Stress field prediction in cantilevered structures using convolutional neural networks. Journal of Computing and Information Science in Engineering, 20(1), 011002.

Oh, S., Jung, Y., Kim, S., Lee, I., & Kang, N. (2019). Deep generative design: Integration of topology optimization and generative models. Journal of Mechanical Design, 141(11).

Oh, S., Jung, Y., Lee, I., & Kang, N. (2018). Design automation by integrating generative adversarial networks and topology optimization. In ASME 2018 International Design Engineering Technical Conferences and Computers and Information in Engineering Conference. American Society of Mechanical Engineers Digital Collection.

Pan, S. J., & Yang, Q. (2009). A survey on transfer learning. IEEE Transactions on knowledge and data engineering, 22(10), 1345-1359.

Qi, C. R., Su, H., Mo, K., & Guibas, L. J. (2017). Pointnet: Deep learning on point sets for 3d classification and segmentation. In Proceedings of the IEEE Conference on Computer Vision and Pattern Recognition (pp. 652-660).

Rawat, W., & Wang, Z. (2017). Deep convolutional neural networks for image classification: A comprehensive review. Neural computation, 29(9), 2352-2449.

Selvaraju, R. R., Cogswell, M., Das, A., Vedantam, R., Parikh, D., & Batra, D. (2017). Grad-cam: Visual explanations from deep networks via gradient-based localization. In Proceedings of the IEEE international conference on computer vision (pp. 618-626).

Shea, K., Aish, R., & Gourtovaia, M. (2005). Towards integrated performance-driven generative design tools.

Automation in Construction, 14(2), 253-264.

Sigmund, O. (2007). Morphology-based black and white filters for topology optimization. Structural and Multidisciplinary Optimization, 33(4-5), 401–424.

Simonyan, K., & Zisserman, A. (2014). Very deep convolutional networks for large-scale image recognition. *arXiv preprint arXiv:1409.1556*.

Singh, V., & Gu, N. (2012). Towards an integrated generative design framework. Design studies, 33(2), 185-207.

Su, H., Maji, S., Kalogerakis, E., & Learned-Miller, E. (2015). Multi-view convolutional neural networks for 3d shape recognition. In Proceedings of the IEEE international conference on computer vision (pp. 945-953).

Sun, H., & Ma, L. (2020). Generative Design by Using Exploration Approaches of Reinforcement Learning in Density-Based Structural Topology Optimization. Designs, 4(2), 10.

Sutradhar, A., Park, J., Haghighi, P., Kresslein, J., Detwiler, D., & Shah, J. J. (2017, August). Incorporating manufacturing constraints in topology optimization methods: A survey. International Design Engineering Technical Conferences and Computers and Information in Engineering Conference (Vol. 58110, p. V001T02A073). American Society of Mechanical Engineers.

Umetani, N. (2017, November). Exploring generative 3D shapes using autoencoder networks. In SIGGRAPH Asia 2017 Technical Briefs (p. 24). ACM.

Umetani, N., & Bickel, B. (2018). Learning three-dimensional flow for interactive aerodynamic design. ACM Transactions on Graphics (TOG), 37(4), 89.

Viana, F. A. (2016). A tutorial on Latin hypercube design of experiments. Quality and reliability engineering international, 32(5), 1975-1985.

Wang, G. G., & Shan, S. (2007). Review of metamodeling techniques in support of engineering design optimization.

Williams, G., Meisel, N. A., Simpson, T. W., & McComb, C. (2019). Design repository effectiveness for 3D convolutional neural networks: application to additive manufacturing (DETC2019-97535). Journal of Mechanical Design, 1-44.

Zhang, Z., Jaiswal, P., & Rai, R. (2018). FeatureNet: Machining feature recognition based on 3D Convolution Neural Network. Computer-Aided Design, 101, 12-22.

Zhou, B., Khosla, A., Lapedriza, A., Oliva, A., & Torralba, A. (2016). Learning deep features for discriminative localization. In Proceedings of the IEEE conference on computer vision and pattern recognition (pp. 2921-2929).

# Appendices

**Appendix A: Data augmentation effect on autoencoder**

The effect of data augmentation was confirmed. We compared the learning curve without data augmentation and the case with data augmentation, as shown in Figure A1. The loss value for data augmentation is relatively small. We checked the example of reconstruction without data augmentation, as shown in Figure A2. These images are the same in Figure 6 but with relatively poor quality.

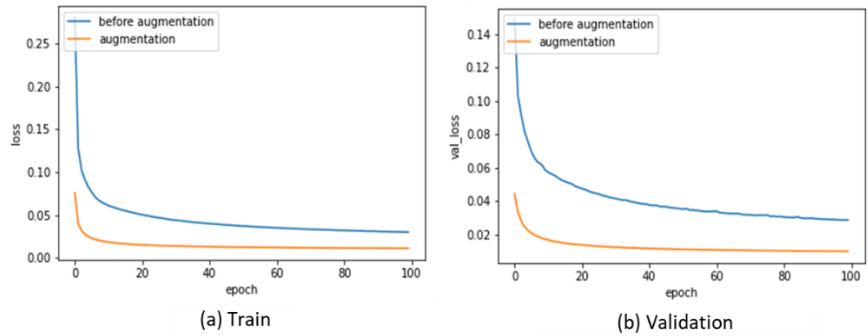

**Figure A1. Comparison of loss value between data augmentation and without data augmentation**

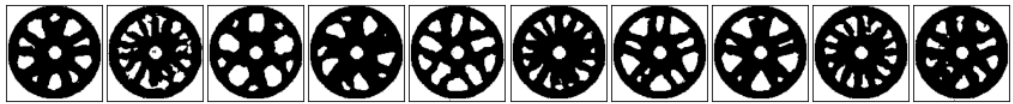

**Figure A2. Reconstructed wheel image without data augmentation**

**Appendix B: Detailed algorithm for sorting and grouping points**

We introduce a detailed algorithm to select and group each point. The following initial preparations are required: The first coordinate $(x0, y0)$ of array $A$, where the whole point was stored, is declared as the initial value. The coordinates $(x0, y0)$ used as the initial value (x_init, y_init) are deleted from array $A$. An array is created to store the points to be grouped. The initial value (x_init, y_init) is stored in the $i$-th array of the group array. At this time, $i$ is zero. At the end of the initial operation, the for loop is executed as follows.

1) The initial value is declared as a fixed point, and the nearest fixed point is obtained in array $A$.

2) The closest point is declared as a new initial value and deleted from array $A$.

3) The distance between the fixed point and the initial value (the closest point to the fixed point) is calculated. If the distance between the fixed point and the initial value does not exceed the threshold, the initial value is stored in the $i$-th group array. Otherwise, a new $(i + 1)$-th group array is created, and the initial value is stored in the $(i + 1)$-th group array. The for loop process is repeated until array $A$ is empty. Any point cannot belong to another group at the same time because it is declared as the initial value and deleted from array $A$.

For determining the distance threshold, an initial test was conducted at equal intervals of 10 steps from 10 to 100. Thus, all points were completely separated into each spoke-shaped group when the threshold reached 10. A second test was conducted at equal intervals of five steps from 1 to 10 to confirm the precise distance threshold. In the second test, all points were completely separated from the threshold of three. Therefore, the final threshold was chosen as five, which is the next value of the lowest threshold. Figure B1 shows the group's separation results for each threshold.

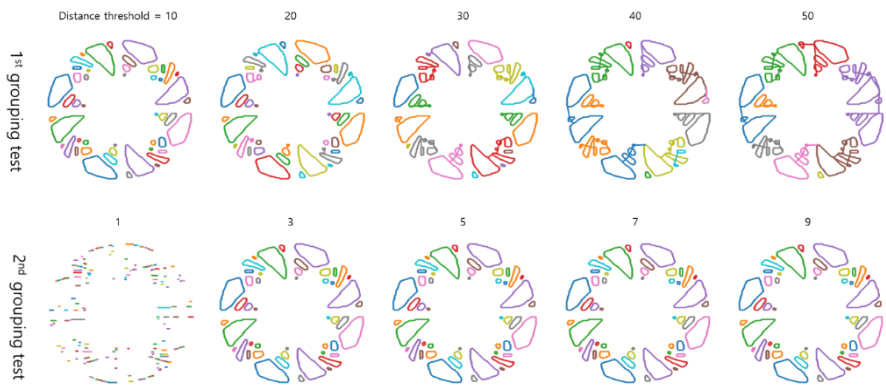

**Figure B1. Tests for determining the distance threshold**

At the end of the above grouping process, all points are sorted in order. We can then take one point of the group array along the same interval and save it as a new array to store fewer points than the existing array. The spline curve becomes a closed curve when the first coordinates of the stored new array are inserted at the end. Closed curves are recognized as surfaces in CAD software, enabling "body" generation. Figure B2 shows examples of spline, reduced point spline, and closed curves.

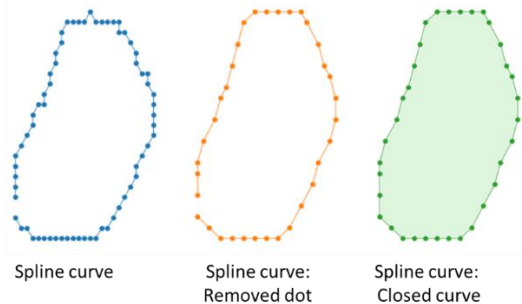

**Figure B2. Example of spline curve**